# Discrete Fresnel Transform and Its Circular Convolution

Xing Ouyang, Cleitus Antony, Fatima Gunning, Hongyu Zhang, and Yong Liang Guan

*Abstract*—Discrete trigonometric transformations, such as the discrete Fourier and cosine/sine transforms, are important in a variety of applications due to their useful properties. For example, one well-known property is the convolution theorem for Fourier transform. In this letter, we derive a discrete Fresnel transform (DFnT) from the infinitely periodic optical gratings, as a linear trigonometric transform. Compared to the previous formulations of DFnT, the DFnT in this letter has no degeneracy, which hinders its mathematic applications, due to destructive interferences. The circular convolution property of the DFnT is studied for the first time. It is proved that the DFnT of a circular convolution of two sequences equals either one circularly convolving with the DFnT of the other. As circular convolution is a fundamental process in discrete systems, the DFnT not only gives the coefficients of the Talbot image, but can also be useful for optical and digital signal processing and numerical evaluation of the Fresnel transform.

## I. INTRODUCTION

LINEAR (integral) transformations play a significant role in signal processing, system modeling and analysis, and other scientific subjects. As a special class of linear transformations, trigonometric transforms

X. Ouyang, C. Antony, F. Gunning, and H. Zhang are with the Photonics Systems Group, Tyndall National Institute, Lee Maltings, Dyke Parade, Cork, Ireland (e-mail: xing.ouyang@tyndall.ie).
Y. L. Guan is with the School of Electrical and Electronic Engineering, Nanyang Technological University, Singapore 639801

and their discrete counterparts are of important due to their unique properties in various applications, such as, digital signal processing (DSP), data compression, communication engineering, and optics and physics. One of the remarkable properties making them useful is their convolution property, i.e., the discrete Fourier transform (DFT) of a circular convolution, and the discrete sine/cosine transform (DST/DCT) of a symmetric convolution [1]. The most well-known example is the convolution theorem which states that, under suitable conditions, the DFT of a circular convolution of two sequences equals the point-wise multiplication of their DFTs.

In this letter, we will study another type of trigonometric transformation, the discrete Fresnel transform (DFnT), which is from the study of Talbot image or Talbot effect in optics [2]. In the literature, the study of DFnT was almost focused on its function to describe the behavior of the near-field diffraction after gratings. In [3-6], the DFnT gives the coefficients of the optical field after the infinitely periodic gratings. However, in the formulation of the DFnT, there exists degeneracy due to destructive interference. As a result, half of the entries in the DFnT matrix becomes zeros if the size of the DFnT matrix is an even number. Such degeneracy hinders its application as a linear transformation, and it is probably why the DFnT is little known by the signal processing society [7]. Moreover, although there exist various formulations for DFnT, the convolution property of the DFnT has not been discussed yet.

In this letter, we will derive the DFnT from the infinitely periodic gratings as a linear transformation. The DFnT derived here also gives the coefficients of the optical field. Depending on the size of the DFnT matrix being even or odd, there are two formulations for the DFnT, and there exists no degeneracy. The properties of the DFnT matrices, such as, their eigenvalues and eigenvectors are studied. Moreover, the circular convolution property of DFnT is investigated for the first time. It is proved that the DFnT of a circular convolution of two sequences equals either one convolving with the DFnT of the other. On the contrary, we will also show that the above circular convolution property does not always hold for the previous DFnT.

## II. Preliminaries and Problem Statement

Fresnel diffraction integrals mathematically formulate the near-field diffraction. When the Fresnel diffraction occurs in the periodic gratings, a self-image is formed, as shown in Fig. 1. The image is called Talbot image or Talbot effect, which was first observed by Talbot in 1836 [8] and first mathematically studied by Rayleigh in 1881 [9]. The Talbot effect is useful in a variety of physical phenomena and applications, such as, laser and waveguide, quantum physics, digital holography and image processing, as well as communication engineering [10-14].

The discrete form of the Fresnel transform, discrete Fresnel transform (DFnT) describes the coefficients of the optical field of the Talbot image. It is critically important for the numerical evaluation of Fresnel diffraction, and has useful applications, as a linear transformation, in optical imaging and DSP, etc.

Various formulations for DFnT were proposed depending on the conditions. For example, in the studies of the Talbot array illuminators, the optical derivations of DFnT was proposed by Szwaykowski, Arrizón, and Castañeda et al [3-6, 15, 16].

In [3], the Talbot coefficients of infinitely periodic gratings at the fractional Talbot distance $2Z_T/N$ is derived, where $Z_T$ is the Talbot length and $2Z_T$ is called the primary Talbot length. The closed-form of the coefficients is given in [4]. Depending on $N \equiv 0, 1, 2$ and $3 \pmod 4$, there are four closed-form expressions for the $N$ coefficients, refer to Eq. (14)-(17) in [4]. In the cases of $N \equiv 0$ and $2 \pmod 4$, there exists degeneracy in the coefficients which is caused by the destructive interference of the diffraction. Half of the $N$ coefficients, as a result, become zeros, as indicated by the cross markers in Fig. 1.

In [5, 6], the coefficients of the Talbot image are given in matrix form. Accordingly, there are four closed-form DFnT matrices. The DFnT matrices have a size of $N$ if $N \equiv 1$ or $3 \pmod 4$, and due to the degeneracy, the size of the matrices reduces to $N/2$ if $N \equiv 0$ or $2 \pmod 4$. Mathematically, such degeneracy hinders the applications of DFnT as a linear transformation, because the size of the DFnT matrices is $N$ if $N$ is odd, and its size shrinks to $N/2$, if $N$ is even.

There are other formulations for DFnT matrix depending on different assumptions [15-19]. In [15, 16], the DFnT is derived from finite periodic grating. In [17-19] and references therein, algebraic approach to DFnT, also called chirp z-transform in the signal processing community, is derived by completing the square of the phase in the DFT. The DFnT matrix in [15-19] has almost the same formulation, see Eq. (17) in [15] and Eq. (12) in [17], and only a single formulation of the DFnT matrix exists without degeneracy.

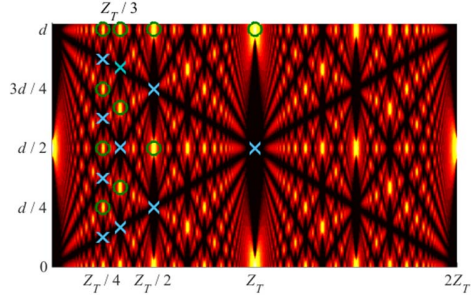

Fig. 1. Example of the optical field of Talbot image at the fractional Talbot distance; the fields at the crosses (×) are zero due to destructive interference and at the circles (○) are nonzero.

Although there were various forms of DFnT in the literature, the convolution property of the DFnT has not been discussed yet. In this letter, we will derive the DFnT from Talbot image without the degeneracy, and it possesses, as mentioned above, a circular convolution property, which does not always holds for the DFnT in the previous formulations.

### III. Continuous Fresnel Transform

In this section, we will first review the continuous Fresnel transform of a single slit (grating), and the DFnT is derived from the infinitely periodic gratings in the following section.

The Fresnel transform occurs in the diffraction after a single slit. Given a monochromatic optical source of wavelength $\lambda$ and a one-dimensional grating whose complex transmittance is $s(t)$ at the plane $z = 0$, the diffractive optical field at some plane $z > 0$, namely the Fresnel transform of $s(t)$ is [2]

$$\hat{s}(\eta) = \mathcal{F}_\alpha\{s(t)\}(\eta) = \frac{1}{\sqrt{\alpha}} e^{-j\frac{\pi}{4}} \int_{-\infty}^{\infty} s(\eta - t) e^{j\frac{\pi}{\alpha}t^2} \, dt, \quad (1)$$

where $\mathcal{F}_\alpha\{\}(\eta)$ denotes the Fresnel transform with $\alpha = \lambda z$.

In the continuous Fresnel transform, F. Gori brought out the convolution property of the Fresnel transform in his chapter 'Why Fresnel transform is so little known?' in [Theorem 1, 7]

$$\hat{x}(\eta) = \mathcal{F}_\alpha \{x(t)\}(\eta) = \mathcal{F}_\alpha \{s_1(t) * s_2(t)\}(\eta) \\ = \hat{s}_1(\eta) * s_2(\eta) = s_1(\eta) * \hat{s}_2(\eta), \quad (2)$$

where $\hat{x}(\eta)$, $\hat{s}_1(\eta)$, and $\hat{s}_2(\eta)$ are the Fresnel transforms of $x(t)$, $s_1(t)$, and $s_2(t)$, respectively. In Eq. (2), the continuous Fresnel transform of a linear convolution of two functions is either one linearly convolving with the Fresnel transform of the other.

## IV. DISCRETE FRESNEL TRANSFORM

In this section, we will arrive at the DFnT from the derivation of the infinitely periodic gratings. The properties of the DFnT matrix in this letter is investigated. Moreover, with the special formulation of the DFnT, the circular convolution property of the DFnT is studied. It is proved that the DFnT of a circular convolution of two sequences is equal to either one circularly convolving with the DFnT of the other.

### A. DFnT from the Talbot Image

Suppose that the one-dimension grating is repeated infinitely and periodically with a distance $d$. Consider the periodic diffraction field within the period from 0 to $d$ at the fractional of the Talbot distance, $z = Z_T/N$ with $Z_T = d^2/\lambda$, as shown in Fig. 1. The one dimensional optical field is

$$\hat{s}_T(\eta) = \frac{1}{\sqrt{\alpha}} e^{-j\frac{\pi}{4}} \int_0^d s(\eta - t) \sum_k e^{j\frac{\pi}{\alpha}(t-kd)^2} \, dt \\ = \frac{1}{\sqrt{d^2/N}} e^{-j\frac{\pi}{4}} \int_0^d s(\eta - t) e^{j\frac{\pi}{d^2/N}t^2} \sum_k e^{-j\frac{2\pi}{d/N}kt} e^{j\pi N k^2} \, dt. \quad (3)$$

Depending on whether $N$ is even or odd, the summation in Eq. (3) can be further given by

$$\sum_k e^{-j\frac{2\pi}{d/N}kt} e^{j\pi N k^2} = \begin{cases} \text{III}_{d/N}(t) & N \equiv 0 \pmod 2 \\ \text{III}_{d/N}\left(t - \frac{1}{2}\frac{d}{N}\right) & N \equiv 1 \pmod 2 \end{cases}, \quad (4)$$

where $\text{III}_{d/N}(t)$ is the Dirac comb function, defined as

$$\text{III}_{d/N}(t) = \frac{d}{N} \sum_k \delta\left(t - k\frac{d}{N}\right). \tag{5}$$

In Eq. (3), the integral interval is within the period from 0 to $d$, and thus the summation is count for $k = 0, 1, \ldots, N - 1$. It can be further deduced as

$$\hat{s}_T(\eta) = \frac{1}{\sqrt{N}} e^{-j\frac{\pi}{4}} \times \begin{cases} \sum_{k=0}^{N-1} s\left(\eta - k\frac{d}{N}\right) e^{j\frac{\pi}{N}k^2} & N \equiv 0 \pmod 2 \\ \sum_{k=0}^{N-1} s\left(\eta - \left(k + \frac{1}{2}\right)\frac{d}{N}\right) e^{j\frac{\pi}{N}\left(k+\frac{1}{2}\right)^2} & N \equiv 1 \pmod 2 \end{cases} \tag{6}$$

As a result, the discrete Talbot coefficients at $\eta = md/N$ if $N \equiv 0 \pmod 2$ or $\eta = (m + 0.5)d/N$ if $N \equiv 1 \pmod 2$ are given by

$$\hat{s}_T(m) = \frac{1}{\sqrt{N}} e^{-j\frac{\pi}{4}} \sum_{n=0}^{N-1} s\left(n\frac{d}{N}\right) \times \begin{cases} e^{j\frac{\pi}{N}(m-n)^2} & N \equiv 0 \pmod 2 \\ e^{j\frac{\pi}{N}\left(m-n+\frac{1}{2}\right)^2} & N \equiv 1 \pmod 2 \end{cases} \tag{7}$$

From Eq. (6) to Eq. (7), due to the circulant property of the Talbot effect, we used the substitution $n = m - k$ to replace the summation index $k$. Eq. (7) is defined as the discrete Fresnel transform of $s(n)$. Accordingly, we formulate the DFnT in this letter by an $N \times N$ matrix $\boldsymbol{\Psi}$ as,

$$\Psi(m, n) = \frac{1}{\sqrt{N}} e^{-j\frac{\pi}{4}} \begin{cases} e^{j\frac{\pi}{N}(m-n)^2} & N \equiv 0 \pmod 2 \\ e^{j\frac{\pi}{N}\left(m-n+\frac{1}{2}\right)^2} & N \equiv 1 \pmod 2 \end{cases} \tag{8}$$

where $\Psi(m, n)$ is the $(m, n)$-th entry of $\boldsymbol{\Psi}$.

The DFnT matrix in Eq. (8) also gives the field coefficients of the Talbot image, as indicated by the circular markers in Fig. 1. As mentioned before, the DFnT in [3-6] is considered at the fraction of the primary Talbot distance $2Z_T / N$. There exists degeneracy if $N$ is even due to the destructive interference, see the cross markers in Fig. 1. Although the self-images is exactly reproduced at the primary Talbot distance, it is also produced at the Talbot distance $Z_T$ with only a half period shift $d/2$. Hence, the field coefficient at the fraction of Talbot length $Z_T / N$ could be sufficient to represent the coefficients.

In this letter, the DFnT gives the close-form coefficients in Eq. (8) depending on even or odd $N$ because the summation in Eq. (4) gives two closed-form expressions in Dirac comb for even and odd $N$. The degeneracy of the coefficients is avoided, which means that all the coefficients, as indicated at the circles in Fig. 1, are nontrivial, no matter what $N$ is. Therefore, the size of the DFnT matrices are always $N$ no matter what $N$ is.

## B. Properties of the DFnT

The DFnT possesses, taking both the even and odd $N$ in Eq. (8), following properties.

*1) Unitary*

That is, $\mathbf{\Psi}^H \mathbf{\Psi} = \mathbf{I}$, where $\mathbf{I}$ is the $N$ by $N$ identity matrix.

*2) Circulant*

The DFnT matrix is circulant for both even and odd $N$. The entries of the DFnT are also periodic, i.e., $\Psi(m, n) = \Psi((m + aN)_N, (n + bN)_N) = \Psi(m + aN, n + bN)$, where $a, b \in \mathbb{Z}$, and $(\cdot)_N$ denotes modulo-$N$. In addition, if $N$ is even, the DFnT matrix $\mathbf{\Psi}$ is symmetrical $\Psi(m, n) = \Psi(n, m)$, while for odd $N$, $\mathbf{\Psi}$ is not symmetrical, and $\Psi(m, n) = \Psi(n - 1, m)$.

This circulant and period properties of the DFnT are crucial for determining the circular convolution property of DFnT. For example, the DFnT defined in [15-19] is not circulant matrix if $N$ is odd. Thus, we can easily prove that the DFnT in [15-19] does not has the circular convolution property for odd $N$.

*3) Eigenvalues and Eigenvectors*

The DFnT can be diagonalized (eigen-decomposited) by the DFT matrix. Suppose that the DFT matrix is normalized, the eigenvalues with respect to the *k*-th column vector of the DFT matrix are

$$\eta_k = \begin{cases} e^{-j\frac{\pi}{N}k^2} & N \equiv 0 \pmod{2} \\ e^{-j\frac{\pi}{N}k(k+1)} & N \equiv 1 \pmod{2} \end{cases} \qquad (9)$$

The eigenvalues and eigenvectors can be easily derived using the property 2 that the DFnT in Eq. (8) are circulant matrices.

*4) Determinant*

The determinants of the DFnT matrices are

$$|\mathbf{\Psi}| = \prod_{k=0}^{N-1}\eta_k = \begin{cases} e^{-j\frac{\pi}{3}(N-1)\left(N-\frac{1}{2}\right)} & N \equiv 0 \pmod{2} \\ e^{-j\frac{\pi}{3}(N^2-1)} & N \equiv 1 \pmod{2} \end{cases}. \qquad (10)$$

*5) Similarity Transformation*

The similarity transformation of a circulant matrix **Z** with respect to the DFnT is itself, i.e.,

$$\hat{\mathbf{Z}} = \mathbf{\Psi}\mathbf{Z}\mathbf{\Psi}^H = \mathbf{Z}, \qquad (11)$$

where the superscript *H* is the Hermitian transpose.

*Proof*: The first column vector of the circulant matrix **Z** is $z = [z(0), z(1), …, z(N-1)]^T$, with the superscript *T* denoting matrix transpose. By polynomial decomposition of a circulant matrix,

$$\mathbf{Z} = \sum_{k=0}^{N-1} z(k)\mathbf{L}^k, \qquad (12)$$

where **L** is an *N* by *N* down-shift matrix with its first column to be $[0, 1, 0, …, 0]^T$, Eq. (11) can be given by

$$\hat{\mathbf{Z}} = \mathbf{\Psi}\mathbf{Z}\mathbf{\Psi}^H = \sum_{k=0}^{N-1} z(k)\mathbf{\Psi}\mathbf{L}^k\mathbf{\Psi}^H. \tag{13}$$

First consider that $N$ is even. If $k = 0$, we have $\mathbf{\Psi}\mathbf{L}^0\mathbf{\Psi}^H = \mathbf{L}^0$. Then, if $k \neq 0$, based on the circular property of $\mathbf{\Psi}$, the $(m, n)$-th entry of $\mathbf{\Psi}\mathbf{L}^k\mathbf{\Psi}^H$ is

$$\sum_{l=0}^{N-1} \Psi(m,l)\Psi^*((l-k)_N, n)$$
$$= \frac{1}{N}\sum_{l=0}^{N-1} e^{j\frac{\pi}{N}(m-l)^2} e^{-j\frac{\pi}{N}(l-k-n)^2} = \delta(m-k-n), \tag{14}$$

where $\delta(k)$ is the Kronecker delta here. From Eq. (14), one can readily deduce that $\mathbf{\Psi}\mathbf{L}^k\mathbf{\Psi}^H = \mathbf{L}^k$, and Eq. (11) becomes

$$\hat{\mathbf{Z}} = \sum_{k=0}^{N-1} z(k)\mathbf{\Psi}\mathbf{L}^k\mathbf{\Psi}^H = \sum_{k=0}^{N-1} z(k)\mathbf{L}^k = \mathbf{Z}. \tag{15}$$

On the other hand, if $N$ is odd, substituting Eq. (8) into (14), and, similarly, one can readily obtain

$$\sum_{l=0}^{N-1} \Psi(m,l)\Psi^*(l-k, n)$$
$$= \frac{1}{N}\sum_{l=0}^{N-1} e^{j\frac{\pi}{N}\left(m-l+\frac{1}{2}\right)^2} e^{-j\frac{\pi}{N}\left(l-k-n+\frac{1}{2}\right)^2} = \delta(m-k-n). \tag{16}$$

Consequently, Eq. (11) holds for both even and odd $N$.

### C. The DFnT of a Circular Convolution

In the following, we will study the DFnT of a circular convolution. Given two length-$N$ sequences, $h(n)$ and $s(n)$, the DFnT of the circular convolution $r(n) = h(n) \circledast s(n)$ is

$$\hat{r}(n) = h(n) \circledast \hat{s}(n) = \hat{h}(n) \circledast s(n), \tag{17}$$

where $\circledast$ denotes circular convolution, and $n = 0, 1, \ldots, N - 1$.

*Proof:* Here we give the circular convolution in matrix form. Define **H** as the *N* by *N* circulant matrix with its first column to be $\boldsymbol{h} = [h(0), h(1), \ldots, h(N-1)]^T$. Similarly, **S** is the circulant matrix with its first column to be $\boldsymbol{s} = [s(0), s(1), \ldots, s(N-1)]^T$. The circular convolution in matrix form is thus

$$\boldsymbol{r} = \mathbf{H}\boldsymbol{s} = \mathbf{S}\boldsymbol{h}. \tag{18}$$

Using the property 1 and 5 above, the DFnT of *r* is

$$\begin{aligned}\hat{\boldsymbol{r}} &= \boldsymbol{\Psi}\boldsymbol{r} = \boldsymbol{\Psi}\mathbf{H}\boldsymbol{\Psi}^H\boldsymbol{\Psi}\boldsymbol{s} = \boldsymbol{\Psi}\mathbf{S}\boldsymbol{\Psi}^H\boldsymbol{\Psi}\boldsymbol{h} \\ &= \mathbf{H}\boldsymbol{\Psi}\boldsymbol{s} = \mathbf{H}\hat{\boldsymbol{s}} \\ &= \mathbf{S}\boldsymbol{\Psi}\boldsymbol{h} = \mathbf{S}\hat{\boldsymbol{h}},\end{aligned} \tag{19}$$

for both even and odd *N*, where $\hat{\boldsymbol{s}} = \boldsymbol{\Psi}\boldsymbol{s}$ and $\hat{\boldsymbol{h}} = \boldsymbol{\Psi}\boldsymbol{h}$ are the DFnT of *s* and *h*, respectively.

It can be observed that the DFnT of a circular convolution in Eq. (17) is the discrete analogue of the continuous Fresnel transform of a linear convolution in Eq. (2). This convolution property is similar to the DFT of a circular convolution and the Fourier transform of a linear convolution. However, in terms of the Fourier transform and DFT, they convert the convolution in one domain to the multiplication in the other domain, and vice versa. On the other hand, both the Fresnel transform and the DFnT in this letter preserves the convolution from one domain to the other domain.

It should be noted that irrespective of whether *N* is even or odd, the formulation of DFnT in [15-19], refer to [Eq. 17, 15], is the same as the DFnT of the even case in Eq. (8). Therefore, the DFnT in [15-19] has the circular convolution property in Eq. (17) if *N* is even. However, if *N* is odd, the DFnT in [15-19] does not possess Property 2 and 5, nor the circular convolution property in Eq. (17).

## V. Conclusion

In this letter, we formulate the DFnT from the Talbot effect as a linear, discrete trigonometric transformation. Compared to the DFnT in previous formulation, the DFnT has no degeneracy and has a circular convolution property. Rather than converting the convolution from one domain to the multiplication in the other domain as other trigonometric transformations dose, the DFnT preserves the convolution in one domain to the other domain. Therefore, with the circular convolution property, the DFnT could be useful for modelling and analyzing discrete systems, such as digital holography, image processing and DSP in which Fresnel Transform applies.

In addition, the DFnT in this letter also gives the coefficients of the Talbot image in a closed form in Eq. (8). It could be the complementary result of the comprehensive work in [3-6, 15, 16] for the design of Talbot illuminator array and for the study of Talbot effect in optics and quantum physics.